# Globalization, Trade, GDP and BRICS from a network analysis perspective

Xiufeng Yan, Qi Tang
Department of Mathematics
Sussex University
Brighton BN1 9QH, UK
Emails: q.tang@sussex.ac.uk, xy33@sussex.ac.uk

**Abstract**: We study the effect of globalization of world economy between 1980 and 2010 by using network analysis technics on trade and GDP data of 71 countries in the world. We draw results distinguishing relatively developing and relatively developed countries during this period of time and point out the standing out economies among the BRICS countries during the years of globalization: within our context of study, China and Russia are the countries that already exhibit developed economy characters, India is next in line but have some unusual features, while Brazil and South Africa still have erratic behaviors.

## 1. Introduction

In this paper, we carry out a network centrality study in relations among import, export and GDP of countries whose data are available through the International Trading Center from 1980 to 2010 (http://www.intracen.org).

We choose the period 1980 to 2010 because during this period, the concept of globalization of the world economy has met its golden years, witnessed by the rise of the BRICS countries. Although terminated with a financial crisis in 2007, its impact is still felt today. From 2010, there is a new trend of re-industrialization in the developed countries. How this trend will play out will only be testified by the data to come in the future years.

First, we look at export-import data: from data analysis point of view, we can place these data into a table, called the International Trade Table here. This is a table showing the transfers of goods between different countries. Each entry of this table indicates the total value of goods exported from one country in the vertical list to the corresponding country in the row. The following table gives a brief example.

|             | Afghanistan | Algeria | Argentina | Australia | Austria |
|-------------|-------------|---------|-----------|-----------|---------|
| Afghanistan | 0           | 0       | 0         | 0         | 0       |
| Algeria     | 0           | 0       | 0         | 32855.59  | 882.02  |
| Argentina   | 0           | 2E+06   | 0         | 621639.8  | 14268   |
| Australia   | 10648.35    | 4496.4  | 282935.97 | 0         | 70948   |
| Austria     | 16721.09    | 273208  | 240317.69 | 953197.5  | 0       |

Figure 1. An example of international Trade Table (Currency unit is $1,000)

From network analysis point of view, the Trade Table can be regarded as a directed and weighted network with nodes representing the countries and edges representing the value of the trade flow. We will call this network the Trade Network in the following.

The mathematical network theory can be used to reveal useful information regarding the structure of the Trade Network. Such analyses focus on the measures and metrics of the Trade Network such as the flow size distribution, degree distribution, centrality and clustering, which will mainly contribute to the intrinsic knowledge of the Trade Network. Bhattacharya et al. [9] studied the scaled link weight distribution of the Trade Network over a period of 53 years and conclude that the link weight has a distribution approximate to a log-normal distribution. They also suggested that the growth of the trade strength with GDP follows a power law. Garlaschelli and Loffredo[4] analyzed the reciprocity structure of the Trade Network and concluded that the time variation of the network link is related to the external factors such as the dynamics of GDP and geopolitical evolution. They pointed out that the topological properties of the Trade Network and dynamics of GDP had intriguing interaction. The fitness model that they used to describe the probability of two vertices in the international network being connected is a function of several fitness variables. They also argue that the GDP of a country could be successfully identified by the corresponding vertex's fitness. Hence, the topology of the network actually influences the GDP of vertices in the network. A previous study by Smith and White[3] applied quantitative network analysis regarding the Trade Network to measure the structure of the world economic system. They analyzed the roles played by particular countries in the global divisions of labor. They argued the fact that some third-world emerging economies are specialized in low-wage manufacturing influences the new trend of international labor flow that is revealed by the evolution of the network structure. Fagiolo[6] applies a residual International Trade Network whose link is depurated from non-network properties such as geographical distance, size and border effects to analyze its topological properties: it was found that, compared with the original Trade Network, the residual network exhibited very different topological structure. The level of clustering of the residual network decreased significantly, which might be explained by the globalization process. Hidalgo et al. [2] formed a network for the products manufactured and exported by countries. They suggested that countries moved towards a clustered sector that represented sophisticated products when they grew economically. Furthermore, they found out that most countries could move to the sector by traversing infrequent distances, suggesting countries with relatively low economic development level had difficulties in manufacturing competitive products for export.

Another stream of literature regarding applications of network theory to Trade Network focuses on the social network models and the economic explanation of their dynamic effects. Jackson[11] gave a brief introduction regarding economic application of social networks and the roles that the social networks played in exchange, learning and diffusion process. Chaney[16] built a dynamic model for the formation of international social network for traders at firm level and used this model to successfully explain the heterogeneous ability of firms to access foreign markets. This model also helped to explain the dynamic evolution of trade flows. Combes et al. [17] show that social networks could help to mitigate the informational barriers in the international trade.

In this paper, we want to explore the relationship between the Trade Network topology and GDP dynamics for countries with different economic development level by examining the relationship between a country's centrality in the network and the country's GDP. We first obtain the centrality weights of relevant countries in the Trade Network over a number of years, and then we look at the relation between this centrality and GDP.

We divide our discussion into four parts. Section 2 introduces the data used. Section 3 introduces the relevant network theory, we concentrate mainly on the mathematical description of adopted centrality measures. Section 4 demonstrates the back testing method and result discussion, we are able to set a classification for countries by their economic development level and then, find characteristic conclusions on their relationship between GDP and Trade Network centralities, we also isolate BRICS countries out and match them to the statistical observations. The last section will summarize the results.

## 2. Data

Between 1980 and 2010, 71 countries' mutual trade data has been collected consistently by the International Trade Center (ITC). All the other remaining countries' data are incomplete. So our data table will include only these 71 countries.

The nominal GDP data of the 71 countries is collected from the World Bank in current US dollar terms. In order to match the time series of the trading data, the time period that the sample covers is also from 1980 to 2010. In order to classify the 71 countries and further explain the statistical results, the following two data sets are also used, i.e., 1) the nominal GDP per capita in current US dollar terms and 2) the Export of Goods and Services as a percentage of GDP. Both indicators are collected from the World Bank. All above data are available online and could be downloaded into excel spreadsheets.

## 3. Methodology

### 3.1 The Trade Network

The matrix for the Trade Network can be presented as

$$T = \begin{bmatrix} f_{11} & \cdots & f_{1n} \\ \vdots & \ddots & \vdots \\ f_{n1} & \cdots & f_{nn} \end{bmatrix}$$

The $f_{ij}$ represents the value of export from country $i$ to country $j$ in terms of current US dollars. It also indicates the value of import of country $j$ from country $i$. Apparently, the above matrix is a square matrix with zero diagonal entries.

In order to exhibit the Trade Network as a standard directed, weighted network, we normalize the entries of the original export-import matrix. Specifically, each entry is redefined as the value of this entry divided by the sum of all the entries. Hence, the adjacency matrix of the Trade Network is:

$$A = \begin{bmatrix} a_{11} & \cdots & a_{1n} \\ \vdots & \ddots & \vdots \\ a_{n1} & \cdots & a_{nn} \end{bmatrix} \qquad \text{where} \quad a_{ij} = \frac{f_{ij}}{\Sigma_{ij} f_{ij}}.$$

## 3.2 Centrality Measures

In network theory, centrality measures the importance of a vertex. In the next three subsections we introduce and discuss three classic measures of centrality according to the nature of Trade Network.

### 3.2.1 Degree Centrality

The degree centrality measure applies the degree of a vertex as the proxy of its centrality. Each vertex in a directed, weighted network has two degree centralities. The out-centrality for vertex i in the Trade Network is defined as

$$K_i^{out} = \sum_j a_{ij}$$

The in-centrality for vertex i in the Trade Network is

$$K_i^{in} = \sum_j a_{ji}$$

Apparently, the out-centrality of country i is simply the proportion of i's export value to the total trade value (The sum of the 71 countries' export values). Meanwhile, the in-centrality of vertex i is the proportion of i's import value to the total trade value. That is, the importance of a country in the network is purely measured by its total export and import.

For Trade Network, the main advantage of degree centrality is that it has a simple and sound economic explanation. Nevertheless, a potential weakness of the degree centrality is that it treats every country intrinsically the same. That is, it measures one country's centrality in the network without considering its trade partners. However, since different countries have significantly different magnitude in the Trade Network it is natural to incorporate the information that reveals the differences of trade partners.

### 3.2.2 Eigenvector Centrality

The eigenvector centrality has the advantage of measuring a vertex's importance in the network by the weights of its edges together with the importance of its neighbors. The mathematical deduction of eigenvector centrality for directed network is displayed below.

Suppose we assign each vertex a centrality value and denote the vector of the initial centrality as $x(0)$. The out-centrality of one vertex is defined to be the sum of the out-centralities of all the other vertices that are pointed by this vertex. Thus,

$$x_i' = \sum_j a_{ij} x_j$$

in matrix form, it is written as

$$x' = Ax$$

Repeat this process t times to reach a proper estimate.

$$x(t) = A^t x(0)$$

Write the initial centrality $x(0)$ as a linear combination of the eigenvectors $v_i$ of the adjacency matrix A. Thus,

$$x(0) = \sum_i c_i v_i$$

Then,

$$x(t) = A^t \sum_i c_i v_i = \sum_i c_i \lambda_i^t v_i = \lambda_1^t \sum_i c_i \frac{\lambda_i^t}{\lambda_1^t} v_i$$

where $\lambda_i$ are the eigenvalues of the adjacency matrix and $\lambda_1$ is the largest eigenvalue. If we repeat this processes for enough times, in which case $t \to \infty$, the dominant term in $x(t)$ is obviously $c_1 \lambda_1^t v_1$. Thus, the limit of proper estimate of out-centrality is actually proportional to $v_1$. Hence, we have the eigenvector out-centrality

$$Ax = \lambda_1 x$$

The eigenvector in-centrality is defined in a similar way. The eigenvector in-centrality of a vertex is assumed to be the sum of the in-centralities of all the other vertices that have edges pointing to this vertex. Thus,

$$x'_i = \sum_j a_{ji} x_j$$

In matrix form, $x' = A^T x$. With a similar reasoning for eigenvector out-centrality, we have

$$A^T x = \lambda_1 x$$

Where $\lambda_1$ is the largest eigenvalue of the matrix $A^T$ (Actually, it is also the largest eigenvalue of A).

For the Trade Network, the eigenvector out-centrality measures a country's export importance in the network by its total export value and the importance of the corresponding export partners. Correspondingly, the eigenvector in-centrality measures a country's import importance in the network by its total import values and the importance of corresponding import partners. Mathematically, in a weighted, directed network, eigenvector centrality measures the importance of vertices by linear combinations of the importance of their neighbors. The coefficients are the weights of corresponding edges.

In the Trade Network, eigenvector centrality allows us to have a more flexible measure for vertices' centrality since it combines the trade value and the importance of corresponding trade partners.

### 3.2.3 Random walk centrality

The random walk centrality also has a comprehensible economic explanation. The mathematical deduction of the random walk centrality for directed network is summarized as follows.

If there is an amount of US dollars starting from a country in the network, at each step it randomly moves along the edges of the Trade Network. If we denote $p_i(t)$ as the probability that the flow of value is at country i at step t, then

$$p_i(t) = \sum_j \frac{a_{ji}}{k_j^{out}} p_j(t-1) \quad (3.1)$$

$$p_i(t-1) = \sum_j \frac{a_{ij}}{k_j^{in}} p_j(t) \quad (3.2)$$

Where $k_j^{out}$ is the out degree of vertex j and $k_j^{in}$ is the in degree of vertex j.

When $t \to \infty$, we have $p = A^T D_{out}^{-1} p$ and $p' = A D_{in}^{-1} p'$ where A is the adjacency matrix of the Trade Network and

$$D_{out}^{-1} = \begin{bmatrix} \frac{1}{k_1^{out}} & 0 & \cdots & 0 & 0 \\ 0 & \frac{1}{k_2^{out}} & \cdots & 0 & 0 \\ \cdots & \cdots & \cdots & \cdots & \cdots \\ 0 & 0 & \cdots & \frac{1}{k_{n-1}^{out}} & 0 \\ 0 & 0 & \cdots & 0 & \frac{1}{k_n^{out}} \end{bmatrix} \quad D_{in}^{-1} = \begin{bmatrix} \frac{1}{k_1^{in}} & 0 & \cdots & 0 & 0 \\ 0 & \frac{1}{k_2^{in}} & \cdots & 0 & 0 \\ \cdots & \cdots & \cdots & \cdots & \cdots \\ 0 & 0 & \cdots & \frac{1}{k_{n-1}^{in}} & 0 \\ 0 & 0 & \cdots & 0 & \frac{1}{k_n^{in}} \end{bmatrix}$$

According to Perron-Frobenius Theorem, the matrix $AD_{in}^{-1}$ and $A^T D_{out}^{-1}$ both have 1 as their largest eigenvalues (They are both Markov matrices since each column of them sum to 1 and all entries are nonnegative). 1 is also the eigenvalue that has an eigenvector with all elements nonnegative.

If we deduce the final probability distribution defined by (3.1), we have $p = A^T D_{out}^{-1} p$. Thus

$$p_1 = \frac{a_{11}}{k_1^{out}} p_1 + \frac{a_{21}}{k_2^{out}} p_2 + \cdots + \frac{a_{n1}}{k_n^{out}} p_n$$

$$\vdots \qquad \vdots$$

$$p_n = \frac{a_{1n}}{k_1^{out}} p_1 + \frac{a_{2n}}{k_2^{out}} p_2 + \cdots + \frac{a_{nn}}{k_n^{out}} p_n$$

Consequently, in the Trade Network, the probability of the flow of value at country i is measured by the sum of the probability that other countries export to i. That is, it is measured by i's import magnitude in the network. Thus we define country i's random walk in-centrality as $p_i$.

Similarly, if we deduce the final probability distribution according to equation (3.2), we have $p' = AD_{in}^{-1} p'$. Thus

$$p_1' = \frac{a_{11}}{k_1^{in}} p_1' + \frac{a_{12}}{k_2^{in}} p_2' + \cdots + \frac{a_{1n}}{k_n^{in}} p_n'$$

$$\vdots \qquad \vdots$$

$$p_n' = \frac{a_{n1}}{k_1^{in}} p_1' + \frac{a_{n2}}{k_2^{in}} p_2' + \cdots + \frac{a_{nn}}{k_n^{in}} p_n'$$

The probability of the flow of value at country i is measured by the sum of the probability that other countries import from i. That is, it is measured by i's export magnitude in the network. Thus we define country i's random walk out-centrality as $p_i'$.

Economically, random walk centrality can be interpreted as the final distribution of the flow of value in the network. The centrality vector also could be interpreted as the vector with elements of the fractions that different countries obtain from the flow. The steady probabilities serve as a measure of countries' centralities in the Trade Network.

## 4. Result Discussion

We mainly discuss 1) the correlation between out-centrality and in-centrality, 2) the centralities and their relationship with corresponding GDP. Our purpose is to find out meaningful statistical relations for economic groups we are interested in.

In order to verify the intuitive assumption that a country's economic magnitude should be correlated with its importance in the Trade Network, we adopt the weighted GDP as the proxy of a country's economic status. Specifically, a country's GDP in a specific year is normalized by the total GDP of all the 71 countries. The selected indicator is the World Bank's GDP figure in current US$. As for the proxies of countries' importance in the Trade Network, we use the 3 centrality measures defined in Section 3. In light of the fact that the 71 countries have significantly different economic development level, we adopt the World Bank's Indicator-GDP per capita in current US$ to divide the 71 countries into 2 different groups. Group 1 has 36 countries that represent the countries with relatively higher GDP per capita. Group 2 consists of 35 countries with relatively lower GDP per capita. Although the classification might be raw and arbitrary, it allows us to compare and analyze the differences of the two groups with respect to correlations between weighted GDP and centralities in the network. The grouping of countries is displayed by Table 1 in Appendix.

**4.1 Data-testing results on correlation between in-centrality and out-centrality**

Since out-centrality measures a country's importance in the Trade Network by its export while in-centrality measures a country's importance in the Trade Network by its import. It is natural to first analyze the correlation between in-centrality and out-centrality to verify whether a country's export status is correlated with its import status.

The confidence level is set as 95% and all calculation is executed using Matlab. The significant rate is defined as the proportion of the number of countries in a group that exhibit significant results in Pearson correlation test to the total number of countries in the group.

The tables that display the specific correlation between in-centrality and out-centrality of the 71 countries are Tables 2, 3 and 4 in Appendix. The result suggests that three different centrality measures have very similar results in correlation tests.

Generally, the statistical results suggest that for the three different centrality measures, most of the 71 countries have significant positive correlation between in-centrality and out-centrality. This result is consistent with the basic theory of international trade that the globalization process is the mutual effect of different economies exercising competitive advantage strategies. If a country's export increases, it suggests that the country uses more resources to produce the export products and then it requires more import to compensate its internal market. Another explanation is that the signal of increased export suggests that the economic development is strengthening which might result in the growth of import.

The significant rates of the two groups of countries with respect to three different centrality measures are displayed below.

| Random Walk Eigenvector Centrality | Group 1 | Group 2 | Total |
|---|---|---|---|
| Significant Rate | 80.56% | 80.00% | 81.70% |

| Direct Eigenvector Centrality | Group 1 | Group 2 | Total |
|---|---|---|---|
| Significant Rate | 94.44% | 80.00% | 87.32% |

| Degree Centrality | Group 1 | Group 2 | Total |
|---|---|---|---|
| Significant Rate | 88.89% | 77.14% | 83.10% |

Figure 2. The significant rate of the two groups with respect to three centrality measures

The statistical results suggest that the countries in Group 1 have higher probability of having significant correlation between in-centrality and out-centrality than countries in Group 2. This indicates that the countries with relatively high economic development level are more likely to have significant correlation between export and import, supporting the theory that the export status and import status of countries with relatively high economic development level are more closely related. It is an indication that countries in Group 1 more effectively exercise the competitive advantage strategy.

With respect to the different statistical results of the three centrality measures, the fact that the eigenvector centrality has the highest significant rate is not surprising since one country's eigenvector centrality is measured by the eigenvector centralities of the other countries. As a result, if most of the countries have significant correlations between in-centrality and out-centrality, this correlation will pass to every strongly connected countries in the network. This suggests that eigenvector centrality is more suitable for describing the interaction in the network.

**4.2 Data-testing results on correlation between GDP and centrality**

In light of the fact that most of the 71 countries' in-centrality and out-centrality have significant correlation, we test the relationship between a country's weighted GDP and centrality in the network by examining the correlation between them. The full statistical result is displayed in Tables 5, 6 and 7 in Appendix.

Generally, most of the 71 countries' weighted GDP exhibit significant linear correlation with centrality. The first significant difference between the two groups of countries is that, compared with the countries of Group 2, the countries of Group 1 have significantly higher probability of exhibiting linear correlation between weighted GDP and out-centrality. The significant rates of the linear relationship between weighted GDP and out-centrality for the two groups is summarized to have a clearer view:

|  | Significant rate of linear relationship between weighted GDP and Out-centrality (Random Walk) | Significant rate of linear relationship between weighted GDP and Out-centrality (Eigenvector Walk) | Significant rate of linear relationship between weighted GDP and Out-centrality (Degree) |
|---|---|---|---|
| Group 1 | 86.11% | 88.89% | 88.89% |
| Group 2 | 68.57% | 68.57% | 71.43% |

Figure 3. The significant rates of linear relationship between GDP and Out-centrality of two groups

The significant differences suggest that the GDP of countries with relatively high economic development level are more likely to be correlated with export status than the GDP of countries with relatively low economic development level. That is, the weighted GDP of countries of Group 1 is more closely related with export status than the GDP of countries of Group 2.

The second result is that, the out-centralities of countries of Group 1 have higher probability of exhibiting significant correlation with GDP than the in-centralities of these countries. In contrast, the in-centralities of countries of Group 2 are more likely to have significant correlation with GDP than the out-centrality of these countries.

| Random Walk Centrality | Regression Model | Group 1 | Group 2 | Total |
|---|---|---|---|---|
|  | In versus GDP | 77.78% | 77.14% | 77.46% |
|  | Out versus GDP | 86.11% | 68.57% | 77.46% |
|  |  |  |  |  |
| Eigenvector Centrality | Regression Model | Group 1 | Group 2 | Total |
|  | In versus GDP | 75.00% | 74.28% | 74.65% |
|  | Out versus GDP | 88.89% | 68.57% | 78.87% |
|  |  |  |  |  |
| Degree Centrality | Regression Model | Group 1 | Group 2 | Total |
|  | In versus GDP | 75.00% | 77.14% | 76.39% |
|  | Out versus GDP | 88.89% | 71.43% | 80.56% |
| The confidence level is 95% for all results | | | | |

Figure 4. The summarized significant rates of correlation between GDP and centrality for two groups

This suggests that the weighted GDP of countries with relatively undeveloped economic level are statistically more closely related to the import status rather than export status. For countries with relatively high economic development level, the weighted GDPs are statistically more closely related to export status rather than import status.

Finally，the significant rates of countries of Group 1 are consistently higher than the significant rates of countries of Group 2 for all the three centrality measures. This suggests that the weighted GDP of relatively developed countries are more closely related to the international trade than the weighted GDP of countries relatively underdeveloped.

From Table 8 in Appendix, we could conclude that the number of countries of Group 1 whose weighted GDP have higher correlation with out-centrality than with in-centrality (the number of countries whose GDP only have significant correlation with out-centrality plus the number of countries whose GDP are significant to both but have higher correlation with out-centrality than in-centrality) is almost identical to the number of countries of Group 1 whose weighted GDP have higher correlation with in-centrality than with out-centrality.

However, in the case of Group 2, the number of countries whose GDP have higher correlation with in-centrality than with out-centrality is significantly greater that the number of countries whose GDP have higher correlation with out-centrality than with in-centrality (18 to 10, 19 to 10 and 20 to 8 for the degree centrality, eigenvector centrality and random walk centrality respectively).

| Degree Centrality | Group 1 | Group 2 |
|---|---|---|
| Correlation with In-centrality greater than Correlation with Out-centrality | 16 | 18 |
| Correlation with Out-centrality greater than Correlation with In-centrality | 16 | 10 |
| | | |
| Eigenvector Centrality | Group 1 | Group 2 |
| Correlation with In-centrality greater than Correlation with Out-centrality | 16 | 19 |
| Correlation with Out-centrality greater than Correlation with In-centrality | 16 | 10 |
| | | |
| Random Walk Centrality | Group 1 | Group 2 |
| Correlation with In-centrality greater than Correlation with Out-centrality | 16 | 20 |
| Correlation with Out-centrality greater than Correlation with In-centrality | 15 | 8 |

Figure 5. The number of countries whose GDP exhibiting significant correlation with centrality in two groups

This result suggests that the relatively underdeveloped Group 2 countries' GDP are indeed more closely related to their import status. In contrast, there is no obvious tendency for relatively developed countries' GDP, which is natural since different countries have different economic development strategies (import-biased growth and export-biased growth) and have different economic structures. The result could be briefly summarized by the table below.

| | Import Centrality | Export Centrality |
|---|---|---|
| Group 1 Countries | Closely Related | Very Closely Related |
| Group 2 Countries | Very Closely Related | Closely Related |

It is also interesting to note that
1) There is almost no country in Group 1 that only exhibits significant correlation between weighted GDP and in-centrality. This might indicate that for countries with relatively high economic development level the status of economic growth is necessarily correlated with the status of export trading.
2) There is almost no country in Group 2 that only exhibits significant correlation between

weighted GDP and out-centrality. This might indicate that for countries with relatively low economic development level, the status of economic growth is necessarily correlated with the status of import trading.

## 4.3 The BRICS countries

In this section, we take a look at the BRICS countries - Brazil, Russia, India, China and South Africa. Since the BRICS countries sustain rapid economic growth and are widely regarded as beneficiaries of the globalization process, the analysis regarding the correlation between their GDP and centralities in the Trade Network may give us some insight into how they fared in the globalization process.

### 4.3.1 The correlation between in-centrality and out-centrality

| Correlation between In-centrality and Out-centrality for BRICS Countries | | | | | | |
|---|---|---|---|---|---|---|
| | Degree Centrality | | Eigenvector Centrality | | Random Walk Centrality | |
| | Correlation | P | Correlation | P | Correlation | P |
| Brazil | 0.221773975 | 0.23051 | -0.229449 | 0.2143599 | 0.0616592 | 0.7417704 |
| Russian Federation | 0.979303341 | 1.1E-21 | 0.9733806 | 4.206E-20 | 0.96718445 | 8.403E-19 |
| India | 0.822227458 | 1.4E-08 | 0.7232935 | 4.287E-06 | 0.81740792 | 2.002E-08 |
| China, P.R.: Mainland | 0.979617054 | 9.1E-22 | 0.9778167 | 3.077E-21 | 0.97879197 | 1.613E-21 |
| South Africa | -0.196766483 | 0.28872 | -0.431558 | 0.015345 | -0.2470896 | 0.180216 |

Figure 6. The correlation between in-centrality and out-centrality of BRICS

The above table suggests:
1) Brazil and South Africa are the two countries in the BRICS group that fail to exhibit significant correlation between the in-centrality and out-centrality. In light of the fact that most of the 71 countries exhibit significant correlation between in-centrality and out-centrality, it could be a signal that the economic structures of the two countries have very unique characteristics.
2) China, India and Russia exhibit very strong correlation between their in-centrality and out-centrality for three different centrality measures, suggesting the export status and import status are very closely related. It indicates that their economic structure and economic growth strategy have characteristics that might be similar to those of countries in Group 1.
3) An additional observation is on India: in the test for eigenvector centrality, the correlation between in-centrality and out-centrality is significantly less than the results from the tests for other two centrality measures. In light of the fact that the eigenvector centrality incorporates the trade value and trade partner's importance, it suggests that India's export and import partners are different countries.

### 4.3.2 The correlation between GDP and centrality

| | Degree Centrality | | | |
|---|---|---|---|---|
| | In Correlation | P | Out Correlation | P |
| Brazil | 0.786132602 | 1.6E-07 | 0.108630567 | 0.56077 |
| Russian Federation | 0.615554154 | 0.00023 | 0.617167772 | 0.00022 |
| India | 0.825560298 | 1.1E-08 | 0.524475007 | 0.00246 |
| China, P.R.: Mainland | 0.94394587 | 1.7E-15 | 0.961136218 | 9.4E-18 |
| South Africa | 0.834893989 | 5.2E-09 | -0.519212482 | 0.00276 |
| | Eigenvector Centrality | | | |
| | In Correlation | P | Out Correlation | P |
| Brazil | 0.782698502 | 2E-07 | -0.008079377 | 0.96559 |
| Russian Federation | 0.651484764 | 7.2E-05 | 0.639934271 | 0.00011 |
| India | 0.882707428 | 5.1E-11 | 0.504298918 | 0.00382 |
| China, P.R.: Mainland | 0.948548604 | 5.1E-16 | 0.945651172 | 1.1E-15 |
| South Africa | 0.842727285 | 2.7E-09 | -0.521741471 | 0.00261 |
| | Eigenvector Centrality | | | |
| | In Correlation | P | Out Correlation | P |
| Brazil | 0.778543713 | 2.5E-07 | 0.167388173 | 0.36809 |
| Russian Federation | 0.614140959 | 0.00024 | 0.610064769 | 0.00027 |
| India | 0.838250654 | 4E-09 | 0.534131445 | 0.00197 |
| China, P.R.: Mainland | 0.951095915 | 2.5E-16 | 0.95470052 | 8.3E-17 |
| South Africa | 0.916683199 | 4.5E-13 | -0.467733615 | 0.00797 |

Figure 7. The correlation between GDP and centrality of BRICS

It can be concluded that
1) China has most significant correlation between GDP growth and import-export status. It is not surprising since China is the second largest economy and largest exporter in the world. Besides, China's economic growth is also well-known to be export-orientated. Therefore the identified significant relationship between China's GDP and centrality in the Trade Network is sound and robust.
2) Russian Federation's GDP also have strong correlations with in-centrality and out-centrality. Mathematically, this could be explained by the strong colinearity of the in-centrality and out-centrality of Russia. Economically, the fact that together with China, the two countries' GDP are highly related with the out-centrality might suggest their economic structure have characteristics that are more similar to countries in Group 1.
3) India's GDP are also significantly correlated with its import status and export status. However, the correlation between India's GDP and out-centrality is significantly less than the correlation between its GDP and in-centrality, which might suggest the economic structure of India is more similar to the countries of Group 2.
4) The correlation between South Africa's GDP and centrality is also significant. However, the negative correlation between its GDP and export-centrality is rather counter intuitive.

Although it may be mathematically explained by the fact that South Africa's in-centrality and out-centrality have significant negative correlation for the measure of eigenvector centrality. Economically, it suggest the economic structure and growth strategy of South Africa is significantly different from the above three countries. Its economic growth is reflected by the increment of import trade. Nevertheless, its export centrality moves towards the opposite direction, which suggests that South Africa's economic growth is not export-biased.

5) The most interesting result is that Brazil's weighted GDP only has significant correlation with its in-centrality. Given the fact that Brazil has one of the fastest GDP growth rates in the world, it hints that the economic development of Brazil have complicated driving factors that can't be simply captured by the network topological properties such as centralities. Meanwhile, the significant correlation between GDP and in-centrality suggests that the economic growth of Brazil is reflected by its growing import trade value.

## 5. Conclusion

This paper analyzes the relationship between a country's weighted GDP and its centrality in the international Trade Network. The sample is divided into two groups according to their economic development level.

Most of the selected 71 countries' weighted GDP have significant correlation with corresponding centrality in the network. Generally, countries with relatively high economic development level have higher probability of presenting significant correlation between centrality and GDP than the countries with relatively low economic development level.

Through the analysis of the two groups, we find out that the probability of relatively developed countries exhibiting significant linear correlation between GDP and out-centrality is significantly higher than the probability of relatively underdeveloped countries presenting such correlation. This result suggests that, compared with the GDP of relatively underdeveloped countries, the GDP of relatively developed countries are more sensitive to the export status. Further analysis suggests that the relatively underdeveloped countries' weighted GDP are more closely related to their import status. In contrast, the relatively developed countries' weighted GDP exhibits no obvious tendency regarding import. This difference might suggest that many of the underdeveloped countries have not been able to effectively exploit their competitive advantage and hence cannot successfully stimulate the economic growth by export. The main conclusion could be seen clearly from Table 9 and Table 10 in the Appendix.

Through the analysis regarding BRICS countries, we conclude that Russia and China's GDP have very strong correlation with their in-centrality and out-centrality. Their GDP is closely related with their export and import status. They are by far the most resembling countries to developed economies among the BRICS.

India's GDP is also significant with both in-centrality and out-centrality. However, India's GDP has a remarkable higher correlation with in-centrality than out-centrality, suggesting that it might have economic structure similar to relatively undeveloped countries. We also noted that India's import and export partners are likely to be very different in terms of importance.

The most counter intuitive result comes from South Africa and Brazil. South Africa's GDP have significant negative correlation with its out-centrality. In contrast, Brazil's GDP is only correlated with its in-centrality. Although there could be plausible economic explanation for the above anomaly, a more reasonable conclusion could be that the Trade Network topology properties may not fully capture the information regarding these countries' economic growth.

Given the fact that different countries have significantly different economic development strategy and economic structure, further research might be carried out on the causality of the preceding conclusion. Meanwhile, the centrality measures could be further improved by adopting metrics that incorporates non-network property for each vertex such as generalized Katz centrality or Page Rank centrality. What's more, HITS centrality measure should also be considered since it ingeniously defines the in-centrality and out-centrality by different types of vertices (This characteristic is admirable for the Trade Network since it could describe the heterogeneity of different countries). Finally, other metrics regarding the Trade Network such as degree distribution or clustering coefficient might also contribute to the cognition of the relationship between GDP and centrality in the Trade Network.

## Appendix: Data Tables

Table 1 Economies grouping

| Economy | GDP per Capita( US$) | Group | Year |
|---|---|---|---|
| Norway | 65188.51794 | 1 | 2014 |
| Qatar | 59893.77807 | 1 | 2014 |
| Switzerland | 59055.41403 | 1 | 2014 |
| Macao SAR, China | 54091.53187 | 1 | 2014 |
| Ireland | 47280.69576 | 1 | 2014 |
| Denmark | 47230.12452 | 1 | 2014 |
| United States | 45710.25267 | 1 | 2014 |
| Sweden | 45587.75844 | 1 | 2014 |
| Netherlands | 42893.242 | 1 | 2014 |
| Austria | 41246.56686 | 1 | 2014 |
| United Kingdom | 40231.02093 | 1 | 2014 |
| Germany | 39219.36712 | 1 | 2014 |
| Finland | 39086.4688 | 1 | 2014 |
| Japan | 37573.36975 | 1 | 2014 |
| Belgium | 37559.4355 | 1 | 2014 |
| Canada | 37524.31519 | 1 | 2014 |
| Australia | 37488.91334 | 1 | 2014 |
| Singapore | 36897.87413 | 1 | 2014 |
| France | 35620.14353 | 1 | 2014 |
| Hong Kong SAR, China | 33534.2779 | 1 | 2014 |
| Kuwait | 31436 | 1 | 2014 |
| Italy | 29408.89507 | 1 | 2014 |
| New Zealand | 29146.11172 | 1 | 2014 |
| United Arab Emirates | 25140.75659 | 1 | 2014 |
| Spain | 25134.35759 | 1 | 2014 |
| Israel | 24340.18602 | 1 | 2014 |
| Korea, Rep. | 23892.5312 | 1 | 2014 |
| Cyprus | 20516.78252 | 1 | 2014 |
| Greece | 18221.50604 | 1 | 2014 |
| Portugal | 18210.14705 | 1 | 2014 |
| Saudi Arabia | 18060.23408 | 1 | 2014 |
| Czech Republic | 14637.50399 | 1 | 2014 |
| Trinidad and Tobago | 14370.17479 | 1 | 2014 |
| Hungary | 11429.85636 | 1 | 2014 |
| Poland | 10781.69667 | 1 | 2014 |
| Chile | 9728.481224 | 1 | 2014 |
|  |  |  |  |
| Turkey | 8722.92227 | 2 | 2014 |
| Mexico | 8519.001632 | 2 | 2014 |
| Malaysia | 6997.671588 | 2 | 2014 |
| Russian Federation | 6923.493388 | 2 | 2014 |
| Venezuela, RB | 6401.905166 | 2 | 2014 |
| Argentina | 6195 | 2 | 2014 |
| Libya | 6125.321419 | 2 | 2014 |
| Romania | 6072.842811 | 2 | 2014 |
| South Africa | 5916.462561 | 2 | 2014 |
| Brazil | 5823.043784 | 2 | 2014 |
| Kazakhstan | 5424.625076 | 2 | 2014 |
| Cuba | 5049 | 2 | 2014 |
| Dominican Republic | 4884.066527 | 2 | 2014 |
| Colombia | 4394.066614 | 2 | 2014 |
| Jamaica | 4189 | 2 | 2014 |
| Peru | 4109.596136 | 2 | 2014 |
| China | 3583.375897 | 2 | 2014 |
| Thailand | 3437.841215 | 2 | 2014 |
| Algeria | 3243.990093 | 2 | 2014 |
| Iran, Islamic Rep. | 3131.798632 | 2 | 2014 |
| Angola | 2737.877052 | 2 | 2014 |
| Iraq | 2517.609508 | 2 | 2014 |
| Ukraine | 2138.275806 | 2 | 2014 |
| Sri Lanka | 2004.255223 | 2 | 2014 |
| Indonesia | 1810.312297 | 2 | 2014 |
| Mongolia | 1795.526752 | 2 | 2014 |
| Philippines | 1581.407551 | 2 | 2014 |
| Egypt, Arab Rep. | 1566.540802 | 2 | 2014 |
| India | 1164.996268 | 2 | 2014 |
| Nigeria | 1055.836611 | 2 | 2014 |
| Vietnam | 1028.623406 | 2 | 2014 |
| Pakistan | 789.5846499 | 2 | 2014 |
| Kenya | 632.3694992 | 2 | 2014 |
| Haiti | 473.2978726 | 2 | 2014 |
| Afghanistan | 415.0034777 | 2 | 2014 |

Table 1 Economies grouping

Table 2 Correlations between eigenvector in-centrality and out-centrality

|  | Correlation | P | Economy Group |
|---|---|---|---|
| Colombia | -0.007658695 | 0.967384 | 2 |
| United Arab Emirates | 0.009356873 | 0.960157 | 1 |
| Argentina | 0.091787979 | 0.623362 | 2 |
| Afghanistan, Islamic Republic of | 0.101719275 | 0.586104 | 2 |
| Israel | 0.132461113 | 0.477491 | 1 |
| Brazil | -0.229449028 | 0.21436 | 2 |
| Sri Lanka | -0.24554419 | 0.183044 | 2 |
| Iraq | 0.256539713 | 0.16359 | 2 |
| Pakistan | -0.26390442 | 0.151415 | 2 |
| Cuba | 0.363730532 | 0.04428 | 2 |
| United States | 0.387704219 | 0.031159 | 1 |
| Romania | 0.397219715 | 0.02692 | 2 |
| South Africa | -0.431558025 | 0.015346 | 2 |
| Finland | 0.490965024 | 0.00504 | 1 |
| Cyprus | 0.502300076 | 0.003983 | 1 |
| Nigeria | 0.553990435 | 0.001223 | 2 |
| Indonesia | 0.560668191 | 0.001035 | 2 |
| Chile | 0.57723756 | 0.000674 | 1 |
| Jamaica | 0.596562289 | 0.000397 | 2 |
| China,P.R.:Macao | -0.599017401 | 0.00037 | 1 |
| Dominican Republic | 0.612868558 | 0.000247 | 2 |
| Norway | 0.618864706 | 0.000206 | 1 |
| Iran, Islamic Republic of | 0.61926218 | 0.000204 | 2 |
| Korea, Republic of | 0.626890141 | 0.000161 | 1 |
| Australia | 0.636438561 | 0.000119 | 1 |
| Kuwait | 0.641548647 | 0.0001 | 1 |
| Portugal | 0.642409863 | 9.76E-05 | 1 |
| Austria | 0.659734791 | 5.41E-05 | 1 |
| Egypt | 0.663295181 | 4.77E-05 | 2 |
| Netherlands | 0.683407579 | 2.26E-05 | 1 |
| Ireland | 0.714049424 | 6.46E-06 | 1 |
| India | 0.723293536 | 4.29E-06 | 2 |
| Saudi Arabia | 0.73236448 | 2.82E-06 | 1 |
| Greece | 0.737594586 | 2.2E-06 | 1 |
| Spain | 0.753673934 | 9.86E-07 | 1 |
| Canada | 0.753962076 | 9.71E-07 | 1 |
| Peru | 0.758353226 | 7.72E-07 | 2 |
| Turkey | 0.767585554 | 4.68E-07 | 2 |
| Haiti | 0.783687519 | 1.85E-07 | 2 |
| Angola | 0.78730345 | 1.49E-07 | 2 |
| Trinidad and Tobago | 0.794627808 | 9.41E-08 | 1 |
| Qatar | 0.799608106 | 6.82E-08 | 1 |
| Italy | 0.804884736 | 4.8E-08 | 1 |
| Venezuela, República Bolivariana de | 0.806378766 | 4.34E-08 | 2 |
| Germany | 0.807934715 | 3.9E-08 | 1 |
| Thailand | 0.808239666 | 3.82E-08 | 2 |
| Kenya | 0.834253704 | 5.52E-09 | 2 |
| France | 0.841061125 | 3.15E-09 | 1 |
| Switzerland | 0.846754589 | 1.93E-09 | 1 |
| Malaysia | 0.855754763 | 8.51E-10 | 2 |
| New Zealand | 0.859176229 | 6.15E-10 | 1 |
| Singapore | 0.864949053 | 3.49E-10 | 1 |
| Sweden | 0.870716747 | 1.92E-10 | 1 |
| Philippines | 0.873135201 | 1.49E-10 | 2 |
| Denmark | 0.875212551 | 1.19E-10 | 1 |
| Algeria | 0.890743515 | 1.92E-11 | 2 |
| Japan | 0.900668103 | 5.14E-12 | 1 |
| Libya | 0.902616251 | 3.91E-12 | 2 |
| Ukraine | 0.908423313 | 1.67E-12 | 2 |
| United Kingdom | 0.91509395 | 5.81E-13 | 1 |
| Mexico | 0.932435683 | 2.37E-14 | 2 |
| China, P.R.: Hong Kong | 0.943685471 | 1.82E-15 | 1 |
| Poland | 0.953205824 | 1.32E-16 | 1 |
| Mongolia | 0.954534479 | 8.76E-17 | 2 |
| Belgium | 0.958108759 | 2.73E-17 | 1 |
| Vietnam | 0.963270002 | 4.2E-18 | 2 |
| Russian Federation | 0.973380632 | 4.21E-20 | 2 |
| Hungary | 0.975289524 | 1.45E-20 | 1 |
| China, P.R.: Mainland | 0.977816735 | 3.08E-21 | 2 |
| Kazakhstan | 0.984918724 | 1.2E-23 | 2 |
| Czech Republic | 0.989376651 | 7.65E-26 | 1 |

Table 3 Correlations between random walk in-centrality and out-centrality

| | Correlation | P | Economy Group |
|---|---|---|---|
| Sri Lanka | 0.040510599 | 0.828696914 | 2 |
| Netherlands | -0.042939766 | 0.818589638 | 1 |
| Brazil | 0.061659199 | 0.74177038 | 2 |
| Israel | -0.093889409 | 0.615397373 | 1 |
| Afghanistan, Islamic Republic of | -0.115862125 | 0.534810604 | 2 |
| Argentina | -0.13102983 | 0.482304152 | 2 |
| Canada | 0.148130923 | 0.426458092 | 1 |
| Austria | 0.165315414 | 0.37413753 | 1 |
| Indonesia | 0.238877649 | 0.195599823 | 2 |
| Finland | 0.241926922 | 0.189784571 | 1 |
| South Africa | -0.247089553 | 0.180216013 | 2 |
| United States | 0.266221657 | 0.147722983 | 1 |
| Dominican Republic | 0.339817567 | 0.061432808 | 2 |
| Pakistan | 0.373183002 | 0.038660651 | 2 |
| United Arab Emirates | 0.379569494 | 0.035200154 | 1 |
| China,P.R.:Macao | 0.41629297 | 0.019836028 | 1 |
| Colombia | 0.420333635 | 0.018553348 | 2 |
| Cuba | 0.449220455 | 0.011240744 | 2 |
| Chile | 0.513244758 | 0.003149537 | 1 |
| Nigeria | 0.53375336 | 0.001985558 | 2 |
| Singapore | 0.544590889 | 0.001537614 | 1 |
| Romania | 0.54521536 | 0.001514733 | 2 |
| Jamaica | 0.545422726 | 0.001507201 | 2 |
| Peru | 0.558908867 | 0.001082284 | 2 |
| Greece | 0.573131663 | 0.000751497 | 1 |
| Portugal | 0.578163153 | 0.000657889 | 1 |
| Germany | 0.588821796 | 0.00049276 | 1 |
| Norway | 0.591854487 | 0.000453026 | 1 |
| Spain | 0.592130684 | 0.000449551 | 1 |
| Australia | 0.648639391 | 7.92739E-05 | 1 |
| Iran, Islamic Republic of | 0.651043266 | 7.30589E-05 | 2 |
| Mexico | 0.654941644 | 6.39022E-05 | 2 |
| Cyprus | -0.660197831 | 5.31848E-05 | 1 |
| Iraq | 0.669039448 | 3.87392E-05 | 2 |
| Denmark | 0.673502136 | 3.28807E-05 | 1 |
| Philippines | 0.702306923 | 1.06352E-05 | 2 |
| Kenya | 0.702907933 | 1.03732E-05 | 2 |
| Egypt | 0.703220024 | 1.02395E-05 | 2 |
| Haiti | 0.703740483 | 1.00199E-05 | 2 |
| France | 0.70610185 | 9.07616E-06 | 1 |
| Italy | 0.711644337 | 7.16839E-06 | 1 |
| Malaysia | 0.712816622 | 6.81467E-06 | 2 |
| New Zealand | 0.734182875 | 2.58958E-06 | 1 |
| Kuwait | 0.754196861 | 9.59606E-07 | 1 |
| Saudi Arabia | 0.756808915 | 8.37217E-07 | 1 |
| Thailand | 0.768942009 | 4.3419E-07 | 2 |
| Angola | 0.774488652 | 3.17372E-07 | 2 |
| Sweden | 0.777968318 | 2.59551E-07 | 1 |
| Switzerland | 0.799964265 | 6.66407E-08 | 1 |
| Venezuela, República Bolivariana de | 0.813114132 | 2.72237E-08 | 2 |
| India | 0.817407917 | 2.00167E-08 | 2 |
| Korea, Republic of | 0.836606768 | 4.55806E-09 | 1 |
| Trinidad and Tobago | 0.836764204 | 4.49961E-09 | 1 |
| Qatar | 0.845116763 | 2.22354E-09 | 1 |
| Turkey | 0.857910732 | 6.94477E-10 | 2 |
| Ireland | 0.867721348 | 2.62879E-10 | 1 |
| Mongolia | 0.869184504 | 2.25929E-10 | 2 |
| China, P.R.: Hong Kong | 0.87750544 | 9.21032E-11 | 1 |
| Libya | 0.882605381 | 5.14335E-11 | 2 |
| Poland | 0.912726735 | 8.52386E-13 | 1 |
| Japan | 0.91958027 | 2.72354E-13 | 1 |
| Algeria | 0.929069795 | 4.69194E-14 | 2 |
| Hungary | 0.930908718 | 3.24445E-14 | 1 |
| United Kingdom | 0.933569289 | 1.86799E-14 | 1 |
| Ukraine | 0.935256724 | 1.30051E-14 | 2 |
| Belgium | 0.940135006 | 4.31078E-15 | 1 |
| Vietnam | 0.943157986 | 2.07379E-15 | 2 |
| Russian Federation | 0.967184451 | 8.40303E-19 | 2 |
| Kazakhstan | 0.976548308 | 6.83611E-21 | 2 |
| China, P.R.: Mainland | 0.978791972 | 1.61344E-21 | 2 |
| Czech Republic | 0.981437512 | 2.37707E-22 | 1 |

Table 4 Correlations between In-degree and out-degree centrality

| | Correlation | P | Economy Group |
|---|---|---|---|
| Netherlands | -0.013635407 | 0.941964 | 1 |
| Afghanistan, Islamic Republic of | 0.095514207 | 0.609269 | 2 |
| Pakistan | 0.12470619 | 0.503867 | 2 |
| Sri Lanka | 0.142496907 | 0.44445 | 2 |
| South Africa | -0.196766483 | 0.288715 | 2 |
| Colombia | 0.203623664 | 0.271899 | 2 |
| Israel | 0.220057296 | 0.234232 | 1 |
| Brazil | 0.221773975 | 0.230511 | 2 |
| China,P.R.:Macao | 0.22487952 | 0.22388 | 1 |
| Argentina | 0.267445518 | 0.1458 | 2 |
| Indonesia | 0.312625679 | 0.086839 | 2 |
| United Arab Emirates | 0.315841448 | 0.083472 | 1 |
| Norway | 0.437983655 | 0.013728 | 1 |
| United States | 0.444354405 | 0.012267 | 1 |
| Romania | 0.47153244 | 0.007411 | 2 |
| Finland | 0.474418098 | 0.007008 | 1 |
| Cuba | 0.478839683 | 0.006427 | 2 |
| Cyprus | -0.495993058 | 0.004545 | 1 |
| Dominican Republic | 0.502922622 | 0.003931 | 2 |
| Nigeria | 0.514055412 | 0.003094 | 2 |
| Austria | 0.519601622 | 0.002738 | 1 |
| Chile | 0.591106111 | 0.000463 | 1 |
| Peru | 0.592870729 | 0.00044 | 2 |
| Singapore | 0.620110285 | 0.000198 | 1 |
| Greece | 0.636679787 | 0.000118 | 1 |
| Iran, Islamic Republic of | 0.686953058 | 1.97E-05 | 2 |
| Iraq | 0.714574873 | 6.31E-06 | 2 |
| Italy | 0.729431667 | 3.24E-06 | 1 |
| Saudi Arabia | 0.740586657 | 1.9E-06 | 1 |
| Germany | 0.7424775 | 1.73E-06 | 1 |
| Ireland | 0.759343769 | 7.32E-07 | 1 |
| Spain | 0.760918662 | 6.73E-07 | 1 |
| Kuwait | 0.763364565 | 5.9E-07 | 1 |
| Egypt | 0.767965742 | 4.58E-07 | 2 |
| Jamaica | 0.775781732 | 2.95E-07 | 2 |
| Australia | 0.779531648 | 2.37E-07 | 1 |
| Portugal | 0.786744037 | 1.54E-07 | 1 |
| Venezuela, República Bolivariana de | 0.789528573 | 1.3E-07 | 2 |
| Angola | 0.793634379 | 1E-07 | 2 |
| Philippines | 0.795327078 | 9E-08 | 2 |
| Kenya | 0.801304636 | 6.1E-08 | 2 |
| Mexico | 0.809604098 | 3.48E-08 | 2 |
| Japan | 0.815010305 | 2.38E-08 | 1 |
| Trinidad and Tobago | 0.815924302 | 2.23E-08 | 1 |
| France | 0.819943741 | 1.66E-08 | 1 |
| India | 0.822227458 | 1.4E-08 | 2 |
| Thailand | 0.831187629 | 7.05E-09 | 2 |
| Qatar | 0.831385015 | 6.94E-09 | 1 |
| Denmark | 0.857922715 | 6.94E-10 | 1 |
| Korea, Republic of | 0.860812501 | 5.25E-10 | 1 |
| Haiti | 0.866044043 | 3.12E-10 | 2 |
| New Zealand | 0.866363119 | 3.02E-10 | 1 |
| Canada | 0.867172584 | 2.78E-10 | 1 |
| China, P.R.: Hong Kong | 0.884229167 | 4.25E-11 | 1 |
| United Kingdom | 0.885084739 | 3.84E-11 | 1 |
| Libya | 0.891023395 | 1.85E-11 | 2 |
| Malaysia | 0.898393582 | 7.03E-12 | 2 |
| Sweden | 0.90306672 | 3.67E-12 | 1 |
| Algeria | 0.903778071 | 3.31E-12 | 2 |
| Turkey | 0.905703825 | 2.5E-12 | 2 |
| Belgium | 0.922967549 | 1.49E-13 | 1 |
| Poland | 0.942328838 | 2.54E-15 | 1 |
| Switzerland | 0.949781962 | 3.59E-16 | 1 |
| Hungary | 0.955521464 | 6.41E-17 | 1 |
| Mongolia | 0.959152718 | 1.91E-17 | 2 |
| Ukraine | 0.971557039 | 1.09E-19 | 2 |
| Vietnam | 0.97478447 | 1.93E-20 | 2 |
| Russian Federation | 0.979303341 | 1.14E-21 | 2 |
| China, P.R.: Mainland | 0.979617054 | 9.12E-22 | 2 |
| Czech Republic | 0.988440245 | 2.59E-25 | 1 |
| Kazakhstan | 0.989372928 | 7.69E-26 | 2 |

Table 5 Correlation between GDP and Random Walk Centrality

| | In Correlation | P | Out Correlation | P | Group |
|---|---|---|---|---|---|
| Sri Lanka | -0.02750059 | 0.883248748 | -0.212450416 | 0.251207 | 2 |
| China,P.R.:Macao | 0.051967045 | 0.781290585 | -0.763807013 | 5.76E-07 | 1 |
| United Kingdom | 0.05666615 | 0.762056214 | -0.063489786 | 0.734375 | 1 |
| Jamaica | 0.110410712 | 0.554325463 | 0.076432194 | 0.682784 | 2 |
| Romania | 0.110411604 | 0.554322243 | -0.263750232 | 0.151663 | 2 |
| Canada | -0.120959879 | 0.516864135 | 0.453390292 | 0.01042 | 1 |
| Ukraine | 0.124687181 | 0.503932136 | 0.040456839 | 0.828921 | 2 |
| Dominican Republic | 0.148754387 | 0.42449215 | 0.442970559 | 0.012572 | 2 |
| Egypt | 0.161063142 | 0.386719888 | 0.326124026 | 0.073377 | 2 |
| Greece | 0.16449441 | 0.376547977 | -0.491583449 | 0.004977 | 1 |
| Philippines | 0.188242936 | 0.310523268 | 0.096815917 | 0.604377 | 2 |
| Iraq | -0.195137321 | 0.292805997 | 0.008604432 | 0.963358 | 2 |
| Norway | -0.208424571 | 0.260511655 | 0.078368203 | 0.675179 | 1 |
| New Zealand | 0.21186845 | 0.252538405 | 0.098819358 | 0.596882 | 1 |
| Netherlands | 0.263829865 | 0.151534845 | 0.198896973 | 0.283421 | 1 |
| Israel | -0.305658659 | 0.094486632 | 0.759133715 | 7.4E-07 | 1 |
| Australia | 0.436987508 | 0.013968842 | 0.43982179 | 0.013292 | 1 |
| Belgium | -0.458047538 | 0.009563298 | -0.454166454 | 0.010273 | 1 |
| Pakistan | 0.466787075 | 0.008115748 | -0.487707976 | 0.005385 | 2 |
| Malaysia | 0.525090472 | 0.00242122 | 0.640000615 | 0.000106 | 2 |
| Kuwait | 0.539761889 | 0.001724958 | 0.814576363 | 2.45E-08 | 1 |
| Mongolia | -0.545848562 | 0.001491835 | -0.46232518 | 0.00883 | 2 |
| Singapore | 0.547293711 | 0.001440704 | 0.949764732 | 3.61E-16 | 1 |
| Vietnam | 0.549324659 | 0.001371431 | 0.573511228 | 0.000744 | 2 |
| Colombia | 0.562597711 | 0.000986131 | 0.105392729 | 0.572575 | 2 |
| Afghanistan, Islamic Republic of | 0.596428668 | 0.00039844 | 0.044611596 | 0.81165 | 2 |
| Russian Federation | 0.614140959 | 0.000237856 | 0.610064769 | 0.000269 | 2 |
| Libya | -0.619502031 | 0.000202225 | -0.600538061 | 0.000354 | 2 |
| Ireland | 0.620775714 | 0.000194492 | 0.585414671 | 0.000541 | 1 |
| Iran, Islamic Republic of | 0.626022992 | 0.000165331 | 0.600477654 | 0.000355 | 2 |
| United States | 0.6312846 | 0.000140064 | 0.437597975 | 0.013821 | 1 |
| Hungary | 0.645084782 | 8.93313E-05 | 0.665613643 | 4.39E-05 | 1 |
| Denmark | 0.657178562 | 5.91251E-05 | 0.574461072 | 0.000726 | 1 |
| China, P.R.: Hong Kong | 0.6588583 | 5.57509E-05 | 0.86244964 | 4.47E-10 | 1 |
| Italy | 0.67347588 | 3.29127E-05 | 0.710466322 | 7.54E-06 | 1 |
| Venezuela, República Bolivariana de | 0.675492043 | 3.05352E-05 | 0.737183349 | 2.24E-06 | 2 |
| Japan | 0.682677662 | 2.32648E-05 | 0.653063952 | 6.82E-05 | 1 |
| Haiti | -0.691255489 | 1.66481E-05 | -0.713038933 | 6.75E-06 | 2 |
| Angola | 0.706377369 | 8.97144E-06 | 0.91577288 | 5.2E-13 | 2 |
| Mexico | 0.706452097 | 8.94323E-06 | 0.555344914 | 0.001183 | 2 |
| United Arab Emirates | 0.714544732 | 6.32197E-06 | 0.779553367 | 2.37E-07 | 1 |
| Poland | 0.726309678 | 3.73734E-06 | 0.73333445 | 2.7E-06 | 1 |
| Argentina | 0.730227211 | 3.11863E-06 | -0.122670169 | 0.51091 | 2 |
| Cyprus | 0.732070984 | 2.86092E-06 | -0.898996686 | 6.48E-12 | 1 |
| France | 0.761911409 | 6.38185E-07 | 0.632830507 | 0.000133 | 1 |
| Peru | 0.764872956 | 5.43482E-07 | 0.668683663 | 3.92E-05 | 2 |
| Austria | 0.767215516 | 4.77847E-07 | 0.420542385 | 0.018489 | 1 |
| Brazil | 0.778543713 | 2.50975E-07 | 0.167388173 | 0.368092 | 2 |
| Saudi Arabia | 0.792858961 | 1.05231E-07 | 0.959433357 | 1.73E-17 | 1 |
| Portugal | 0.796982382 | 8.08999E-08 | 0.41289245 | 0.020972 | 1 |
| Chile | 0.808653651 | 3.71656E-08 | 0.67437109 | 3.18E-05 | 1 |
| Nigeria | 0.811142687 | 3.12706E-08 | 0.451296729 | 0.010825 | 2 |
| Switzerland | 0.819239628 | 1.75129E-08 | 0.873920526 | 1.37E-10 | 1 |
| Indonesia | 0.81973928 | 1.68817E-08 | 0.443268898 | 0.012506 | 2 |
| Spain | 0.822148462 | 1.41207E-08 | 0.392749384 | 0.028849 | 1 |
| India | 0.838250654 | 3.98069E-09 | 0.534131445 | 0.001968 | 2 |
| Germany | 0.840538151 | 3.28863E-09 | 0.525706397 | 0.002388 | 1 |
| South Africa | 0.842727285 | 2.73161E-09 | -0.521741471 | 0.002611 | 2 |
| Kazakhstan | 0.868518932 | 2.42099E-10 | 0.838981358 | 3.75E-09 | 2 |
| Finland | 0.869279969 | 2.23693E-10 | 0.196101095 | 0.290382 | 1 |
| Sweden | 0.871874114 | 1.70204E-10 | 0.654608853 | 6.46E-05 | 1 |
| Algeria | 0.883721263 | 4.51158E-11 | 0.816504338 | 2.14E-08 | 2 |
| Kenya | 0.897397012 | 8.05008E-12 | 0.61310874 | 0.000245 | 2 |
| Cuba | 0.898388214 | 7.03908E-12 | 0.64777352 | 8.16E-05 | 2 |
| Turkey | 0.905188687 | 2.69645E-12 | 0.751880784 | 1.08E-06 | 2 |
| Korea, Republic of | 0.919472598 | 2.77494E-13 | 0.859040187 | 6.23E-10 | 1 |
| Qatar | 0.925875906 | 8.70286E-14 | 0.845340565 | 2.18E-09 | 1 |
| Thailand | 0.926069394 | 8.39979E-14 | 0.669023561 | 3.88E-05 | 2 |
| Trinidad and Tobago | 0.930461869 | 3.552E-14 | 0.861496746 | 4.91E-10 | 1 |
| Czech Republic | 0.936905595 | 9.04298E-15 | 0.930294484 | 3.67E-14 | 1 |
| China, P.R.: Mainland | 0.951095915 | 2.4652E-16 | 0.95470052 | 8.31E-17 | 2 |

Table 6 Correlation between GDP and Eigenvector Centrality

| | In Correlation | P | Out Correlation | P | Group |
|---|---|---|---|---|---|
| Sri Lanka | -0.061965189 | 0.740533 | -0.161701034 | 0.38482 | 2 |
| United Kingdom | 0.077375067 | 0.679076 | 0.038331783 | 0.83779 | 1 |
| Jamaica | 0.078771633 | 0.673599 | -0.017072059 | 0.92737 | 2 |
| Egypt | 0.095815911 | 0.608133 | 0.158904214 | 0.3932 | 2 |
| Norway | -0.125739491 | 0.500311 | 0.39591613 | 0.02747 | 1 |
| New Zealand | 0.136496032 | 0.464056 | 0.184036418 | 0.32166 | 1 |
| Dominican Republic | 0.138101243 | 0.458767 | 0.453764972 | 0.01035 | 2 |
| Greece | 0.142443669 | 0.444622 | -0.416345134 | 0.01982 | 1 |
| Israel | -0.142654504 | 0.443941 | 0.766964089 | 4.8E-07 | 1 |
| Cyprus | -0.181847816 | 0.327544 | -0.83158855 | 6.8E-09 | 1 |
| Ukraine | 0.189861357 | 0.306305 | 0.095839251 | 0.60805 | 2 |
| Iraq | -0.193088091 | 0.298004 | 0.482579794 | 0.00597 | 2 |
| Philippines | 0.238980826 | 0.195401 | 0.184295883 | 0.32096 | 2 |
| Portugal | 0.24975656 | 0.175408 | 0.29270276 | 0.11005 | 1 |
| Canada | -0.292102425 | 0.110819 | 0.069384096 | 0.71072 | 1 |
| Australia | 0.321519747 | 0.077774 | 0.63257689 | 0.00013 | 1 |
| Mongolia | -0.345261045 | 0.057132 | -0.266702806 | 0.14696 | 2 |
| Pakistan | 0.351652849 | 0.052388 | -0.855023187 | 9.1E-10 | 2 |
| Romania | 0.379095815 | 0.035448 | -0.150378851 | 0.41939 | 2 |
| Ireland | 0.385801991 | 0.032069 | 0.706059991 | 9.1E-06 | 1 |
| China, P.R.: Hong Kong | 0.414274715 | 0.020504 | 0.661609766 | 5.1E-05 | 1 |
| Netherlands | 0.425023951 | 0.017151 | 0.589336939 | 0.00049 | 1 |
| Belgium | -0.440233702 | 0.013196 | -0.419870043 | 0.0187 | 1 |
| Iran, Islamic Republic of | 0.471567485 | 0.007406 | 0.533280772 | 0.00201 | 2 |
| Kuwait | 0.511667518 | 0.003259 | 0.852356729 | 1.2E-09 | 1 |
| Malaysia | 0.539112002 | 0.001752 | 0.665049841 | 4.5E-05 | 2 |
| Afghanistan, Islamic Republic of | 0.543574399 | 0.001576 | 0.193242915 | 0.29761 | 2 |
| Vietnam | 0.574380669 | 0.000727 | 0.586210272 | 0.00053 | 2 |
| Colombia | 0.582646924 | 0.000583 | -0.158894783 | 0.39323 | 2 |
| Libya | -0.605937342 | 0.000303 | -0.560107727 | 0.00105 | 2 |
| Venezuela, República Bolivariana d | 0.642739029 | 9.66E-05 | 0.690811807 | 1.7E-05 | 2 |
| United States | 0.651069047 | 7.3E-05 | 0.508023915 | 0.00353 | 1 |
| Russian Federation | 0.651484764 | 7.2E-05 | 0.639934271 | 0.00011 | 2 |
| Singapore | 0.652078035 | 7.05E-05 | 0.868796742 | 2.4E-10 | 1 |
| Argentina | 0.660345459 | 5.29E-05 | -0.236429423 | 0.20036 | 2 |
| Denmark | 0.674969856 | 3.11E-05 | 0.719681312 | 5E-06 | 1 |
| Germany | 0.68936758 | 1.79E-05 | 0.560294698 | 0.00105 | 1 |
| Mexico | 0.690535393 | 1.71E-05 | 0.606073599 | 0.0003 | 2 |
| Poland | 0.697583525 | 1.29E-05 | 0.710977829 | 7.4E-06 | 1 |
| Hungary | 0.706876484 | 8.78E-06 | 0.650925083 | 7.3E-05 | 1 |
| Angola | 0.713468052 | 6.63E-06 | 0.91361001 | 7.4E-13 | 2 |
| Austria | 0.729114003 | 3.28E-06 | 0.611292605 | 0.00026 | 1 |
| Japan | 0.745300371 | 1.51E-06 | 0.696255707 | 1.4E-05 | 1 |
| Indonesia | 0.749996226 | 1.19E-06 | 0.402862001 | 0.02464 | 2 |
| Korea, Republic of | 0.754534199 | 9.43E-07 | 0.504677718 | 0.00379 | 1 |
| Peru | 0.7705189 | 3.98E-07 | 0.725172452 | 3.9E-06 | 2 |
| United Arab Emirates | 0.780341803 | 2.26E-07 | 0.44282907 | 0.0126 | 1 |
| Haiti | -0.781485534 | 2.11E-07 | -0.668770191 | 3.9E-05 | 2 |
| Brazil | 0.782698502 | 1.96E-07 | -0.008079377 | 0.96559 | 2 |
| Saudi Arabia | 0.789154909 | 1.33E-07 | 0.950418843 | 3E-16 | 1 |
| Italy | 0.791470121 | 1.15E-07 | 0.843112676 | 2.6E-09 | 1 |
| Kenya | 0.81427267 | 2.51E-08 | 0.700360921 | 1.2E-05 | 2 |
| Nigeria | 0.818929714 | 1.79E-08 | 0.487689625 | 0.00539 | 2 |
| Chile | 0.82137469 | 1.5E-08 | 0.65257662 | 6.9E-05 | 1 |
| Turkey | 0.827791517 | 9.2E-09 | 0.637590852 | 0.00011 | 2 |
| Finland | 0.833035947 | 6.09E-09 | 0.623088872 | 0.00018 | 1 |
| Switzerland | 0.834499171 | 5.41E-09 | 0.906244214 | 2.3E-12 | 1 |
| Algeria | 0.842502872 | 2.78E-09 | 0.709315235 | 7.9E-06 | 2 |
| Spain | 0.849193893 | 1.55E-09 | 0.589874216 | 0.00048 | 1 |
| France | 0.861162111 | 5.07E-10 | 0.819420168 | 1.7E-08 | 1 |
| Cuba | 0.862305652 | 4.54E-10 | 0.505186647 | 0.00375 | 2 |
| India | 0.882707428 | 5.08E-11 | 0.504298918 | 0.00382 | 2 |
| Kazakhstan | 0.886148662 | 3.38E-11 | 0.880106631 | 6.9E-11 | 2 |
| Trinidad and Tobago | 0.905090476 | 2.74E-12 | 0.885391612 | 3.7E-11 | 1 |
| Sweden | 0.913290498 | 7.79E-13 | 0.81309496 | 2.7E-08 | 1 |
| South Africa | 0.916683199 | 4.46E-13 | -0.467733615 | 0.00797 | 2 |
| Qatar | 0.918590947 | 3.23E-13 | 0.756077721 | 8.7E-07 | 1 |
| Thailand | 0.935880067 | 1.13E-14 | 0.768967305 | 4.3E-07 | 2 |
| Czech Republic | 0.938946257 | 5.69E-15 | 0.924636194 | 1.1E-13 | 1 |
| China, P.R.: Mainland | 0.948548604 | 5.06E-16 | 0.945651172 | 1.1E-15 | 2 |
| China,P.R.:Macao | 0.952909535 | 1.44E-16 | -0.626587907 | 0.00016 | 1 |

Table 7 Correlation between GDP and Degree Centrality

| | In Correlation | P | Out Correlation | P | Group |
|---|---|---|---|---|---|
| United Kingdom | 0.027729004 | 0.882286 | -0.040208935 | 0.829954 | 1 |
| Sri Lanka | -0.038214869 | 0.838274 | -0.128074276 | 0.492322 | 2 |
| Greece | 0.051177866 | 0.784534 | -0.485454334 | 0.005635 | 1 |
| Ukraine | 0.111245729 | 0.551315 | 0.058597515 | 0.75419 | 2 |
| Israel | 0.137616612 | 0.460361 | 0.82609533 | 1.05E-08 | 1 |
| Canada | 0.154627781 | 0.40622 | 0.37079119 | 0.040025 | 1 |
| Egypt | 0.164867934 | 0.37545 | 0.243011621 | 0.187745 | 2 |
| Dominican Republic | 0.17329749 | 0.351177 | 0.506551198 | 0.003639 | 2 |
| New Zealand | 0.176034488 | 0.343503 | 0.166650365 | 0.370238 | 1 |
| Jamaica | 0.208076566 | 0.261326 | -0.015003929 | 0.93615 | 2 |
| Iraq | -0.22120757 | 0.231734 | 0.089755871 | 0.631103 | 2 |
| China,P.R.:Macao | 0.281614164 | 0.124838 | -0.765315012 | 5.31E-07 | 1 |
| Philippines | 0.292621819 | 0.110157 | 0.076013657 | 0.684432 | 2 |
| Norway | -0.297427886 | 0.104169 | 0.333442849 | 0.066788 | 1 |
| Romania | 0.301558584 | 0.09922 | -0.327328918 | 0.072259 | 2 |
| Netherlands | 0.306771699 | 0.093232 | 0.27198183 | 0.138829 | 1 |
| Ireland | 0.334363615 | 0.065993 | 0.655118024 | 6.35E-05 | 1 |
| Japan | 0.447186986 | 0.01166 | 0.704371343 | 9.76E-06 | 1 |
| Belgium | -0.45633352 | 0.009871 | -0.450586817 | 0.010966 | 1 |
| Singapore | 0.471124675 | 0.007469 | 0.928757297 | 4.99E-14 | 1 |
| Pakistan | 0.471658346 | 0.007393 | -0.65584426 | 6.19E-05 | 2 |
| Colombia | 0.474489311 | 0.006998 | 0.196672498 | 0.28895 | 2 |
| Australia | 0.512275137 | 0.003217 | 0.578533219 | 0.000651 | 1 |
| Kuwait | 0.517052862 | 0.002897 | 0.829348682 | 8.15E-09 | 1 |
| Afghanistan, Islamic Republic of | 0.526487832 | 0.002346 | 0.074063629 | 0.692129 | 2 |
| Cyprus | 0.546856769 | 0.001456 | -0.908246017 | 1.71E-12 | 1 |
| Mongolia | -0.552026184 | 0.001284 | -0.513852039 | 0.003108 | 2 |
| Vietnam | 0.568465684 | 0.000849 | 0.586025013 | 0.000532 | 2 |
| Malaysia | 0.572322837 | 0.000768 | 0.650648471 | 7.4E-05 | 2 |
| Hungary | 0.608249435 | 0.000283 | 0.688944453 | 1.82E-05 | 1 |
| Iran, Islamic Republic of | 0.612146649 | 0.000252 | 0.60059934 | 0.000354 | 2 |
| Russian Federation | 0.615554154 | 0.000228 | 0.617167772 | 0.000217 | 2 |
| China, P.R.: Hong Kong | 0.653383621 | 6.74E-05 | 0.862596235 | 4.41E-10 | 1 |
| Libya | -0.663347947 | 4.76E-05 | -0.581811767 | 0.000597 | 2 |
| Denmark | 0.665923292 | 4.34E-05 | 0.627587148 | 0.000157 | 1 |
| United Arab Emirates | 0.669229246 | 3.85E-05 | 0.780121776 | 2.29E-07 | 1 |
| Argentina | 0.681730156 | 2.41E-05 | 0.362247979 | 0.045217 | 2 |
| Venezuela, República Bolivariana de | 0.682254932 | 2.36E-05 | 0.752865671 | 1.03E-06 | 2 |
| Italy | 0.709418203 | 7.89E-06 | 0.756991122 | 8.29E-07 | 1 |
| Angola | 0.746474179 | 1.42E-06 | 0.92126229 | 2.03E-13 | 2 |
| Austria | 0.74665595 | 1.41E-06 | 0.521893035 | 0.002602 | 1 |
| Portugal | 0.755124489 | 9.14E-07 | 0.478688297 | 0.006446 | 1 |
| Mexico | 0.758106362 | 7.82E-07 | 0.638538328 | 0.000111 | 2 |
| United States | 0.762524395 | 6.17E-07 | 0.39138442 | 0.02946 | 1 |
| Peru | 0.76528392 | 5.31E-07 | 0.704012353 | 9.91E-06 | 2 |
| France | 0.767459184 | 4.71E-07 | 0.679557811 | 2.62E-05 | 1 |
| Poland | 0.773207904 | 3.41E-07 | 0.697408953 | 1.3E-05 | 1 |
| Brazil | 0.786132602 | 1.6E-07 | 0.108630567 | 0.560769 | 2 |
| Spain | 0.786945154 | 1.52E-07 | 0.516377367 | 0.002941 | 1 |
| Saudi Arabia | 0.790026194 | 1.26E-07 | 0.951900561 | 1.95E-16 | 1 |
| Finland | 0.794382032 | 9.56E-08 | 0.403749088 | 0.024293 | 1 |
| Nigeria | 0.802561943 | 5.61E-08 | 0.476336627 | 0.006751 | 2 |
| Haiti | -0.811720627 | 3E-08 | -0.698954186 | 1.22E-05 | 2 |
| Chile | 0.825153384 | 1.13E-08 | 0.717707553 | 5.5E-06 | 1 |
| India | 0.825560298 | 1.09E-08 | 0.524475007 | 0.002455 | 2 |
| Indonesia | 0.834788451 | 5.29E-09 | 0.533088222 | 0.002016 | 2 |
| South Africa | 0.834893989 | 5.24E-09 | -0.519212482 | 0.002762 | 2 |
| Sweden | 0.857643581 | 7.12E-10 | 0.735094815 | 2.48E-06 | 1 |
| Germany | 0.859045552 | 6.23E-10 | 0.583910909 | 0.000564 | 1 |
| Kazakhstan | 0.866412743 | 3.01E-10 | 0.859379755 | 6.03E-10 | 2 |
| Switzerland | 0.867945895 | 2.57E-10 | 0.90254644 | 3.95E-12 | 1 |
| Kenya | 0.874095748 | 1.34E-10 | 0.63124267 | 0.00014 | 2 |
| Turkey | 0.886416302 | 3.27E-11 | 0.785568598 | 1.65E-07 | 2 |
| Cuba | 0.903055495 | 3.67E-12 | 0.642032193 | 9.89E-05 | 2 |
| Algeria | 0.905219409 | 2.68E-12 | 0.816493864 | 2.14E-08 | 1 |
| Korea, Republic of | 0.913751913 | 7.23E-13 | 0.855156684 | 9.01E-10 | 1 |
| Qatar | 0.913774474 | 7.2E-13 | 0.785497055 | 1.66E-07 | 1 |
| Thailand | 0.917735378 | 3.74E-13 | 0.709662274 | 7.8E-06 | 2 |
| Trinidad and Tobago | 0.924480333 | 1.13E-13 | 0.889768745 | 2.16E-11 | 1 |
| Czech Republic | 0.937814553 | 7.37E-15 | 0.932430248 | 2.37E-14 | 1 |
| China, P.R.: Mainland | 0.94394587 | 1.7E-15 | 0.961136218 | 9.39E-18 | 2 |

Table 8 Distribution regarding correlations between centrality and GDP

| Degree Centrality | | |
|---|---|---|
| | Group 1 | Group 2 |
| GDP only significant with In-centrality | 0 | 3 |
| GDP only significant with out-centrality | 5 | 1 |
| Correlation with in-centrality greater than Correlation with out-centrality | 16 | 15 |
| Correlation with out-centrality greater than Correlation with in-centrality | 11 | 9 |
| Non-significant to both | 4 | 7 |
| Eigenvector Centrality | | |
| | Group 1 | Group 2 |
| GDP only significant with In-centrality | 0 | 5 |
| GDP only significant with out-centrality | 5 | 3 |
| Correlation with in-centrality greater than Correlation with out-centrality | 16 | 14 |
| Correlation with out-centrality greater than Correlation with in-centrality | 11 | 7 |
| Non-significant to both | 4 | 6 |
| Random walk Centrality | | |
| | Group 1 | Group 2 |
| GDP only significant with In-centrality | 1 | 6 |
| GDP only significant with out-centrality | 5 | 1 |
| Correlation with in-centrality greater than Correlation with out-centrality | 16 | 14 |
| Correlation with out-centrality greater than Correlation with in-centrality | 10 | 7 |
| Non-significant to both | 4 | 7 |

Table 9: Average significant rate of three centrality measures

| | In-Centrality | Out-Centrality |
|---|---|---|
| GDP of Group 1 | 75.93% | 87.96% |
| GDP of Group 2 | 76.19% | 69.52% |

Three centrality measures' average significant rates for Correlation between GDP and Centrality

Table 10: Correlation Tendency

| | In-correlation greater than Out-correlation | In-correlation less than Out-correlation |
|---|---|---|
| Countries of Group 1 | 16 | 16 |
| Countries of Group 2 | 20 | 10 |

In-correlation : The correlation between In-centrality and GDP
Out-correlation : The correlation between Out-centrality and GDP
The number of countries is averaged from the result for three centraity measures